\documentclass[galaxies,article,accept,pdftex,moreauthors]{Definitions/mdpi} 
\usepackage[utf8]{inputenc}
\usepackage{aas_macros}
\firstpage{1} 
\pubvolume{1}
\issuenum{1}
\articlenumber{0}
\pubyear{2022}
\copyrightyear{2022}
\externaleditor{Academic Editor:Francesca Loi, Tiziana Venturi, Fulai Guo 
 }
\datereceived{08 November 2021} 
\dateaccepted{30 December 2021} 
\datepublished{} 
\hreflink{https://doi.org/} 

\Title{On the Polarisation of Radio Relics}

\TitleCitation{On the Polarisation of Radio Relics}

\Author{Matthias Hoeft $^{1}$\orcidA, Kamlesh Rajpurohit $^{1,2,3}$, Denis Wittor $^{2,4}$, Gabriella di Gennaro $^{4,5}$ \mbox{and Paola Dom\'inguez-Fern\'andez $^{4,6}$}}

\AuthorNames{Matthias Hoeft, Kamlesh Rajpurohit, Denis Wittor, Gabriella di Gennaro and Paola Dom\'inguez-Fern\'andez}

\AuthorCitation{Hoeft, M.; Rajpurohit, K.; Wittor, D.; di Gennaro, G.; Dom\'inguez-Fern\'andez, P.}

\address{%
$^{1}$ \quad Th\"uringer Landessternwarte, Sternwarte 5, 07778 Tautenburg, Germany \\
$^{2}$ \quad Dipartimento di Fisica e Astronomia, Universit\`a di Bologna, Via Gobetti 92/3, 40129 Bologna, Italy \\
$^{3}$ \quad Institute of Radioastronomy --
 INAF, Via Gobetti 101, 40129 Bologna, Italy \\
$^{4}$ \quad Hamburger Sternwarte, Gojenbergsweg 112, 21029 Hamburg, Germany \\
$^{5}$ \quad Leiden Observatory, Leiden University, PO Box 9513, 2300 RA Leiden, The Netherlands \\
$^{6}$ \quad  Department of Physics, School of Natural Sciences, 
Ulsan National Institute of Science and Technology, \mbox{Ulsan 44919,  Korea}
}

\corres{Correspondence: hoeft@tls-tautenburg.de }

\abstract{
  Radio relics are extended radio emission features which trace shock waves in the periphery of galaxy clusters originating from cluster mergers. Some radio relics show a highly polarised emission, which make relics an excellent probe for the magnetisation of the intra-cluster medium. The origin of the relic polarisation is still debated. It could be a result of tangentially stretching the magnetic field at the shock surface. This scenario would naturally explain the alignment of the polarisation (E-vectors) with the shock normal. We have implemented a toy model for the relic polarisation according to this scenario. We find that the magnetic field strength itself crucially affects the fractional polarisation. Moreover, we find that the shock strength has surprisingly little effect on the overall polarisation fraction. Finally, we find that the fractional polarisation may decrease downstream depending on the magnetic field strength. Our results demonstrates that the shock compression scenario provides a very plausible explanation for the radio relic polarisation which specific features permitting to test the origin of radio relic polarisation.
  }


\begin{document}

\section{Introduction}
\label{sec::intro}

 %
 When galaxy clusters merge, large shock waves form in the intra-cluster medium. These shock fronts have been observed as discontinuities in the X-ray surface brightness. Moreover, large-scale radio emission features have been found in the periphery of galaxy clusters and are classified as radio relics~\cite{2019SSRv..215...16V}. Their existence is attributed to the acceleration of thermal or mildly relativistic electrons to higher relativistic energies at the shock fronts~\cite{1998A&A...332..395E,2007MNRAS.375...77H}. Evidence for the connection of radio relics to the merger shocks especially comes from the fact that for several clusters a discontinuity in X-ray has been found to coincide with the radio relic. The radio relic emission manifests that synchrotron radiation is emitted at the shock fronts, hence magnetic fields and electrons with relativistic energies (cosmic ray electrons, CRe) are present in the outskirts of galaxy clusters. A plausible explanation for the origin of CRe is diffusive shock acceleration (DSA)~\cite{1987PhR...154....1B}, since, this mechanism naturally explains the power-law integrated radio spectra found for the radio relics and their dependence on the shock strength~\cite{2007MNRAS.375...77H}. However, a substantial fraction of the energy dissipated at the shock front would need to be channelled into the acceleration of electrons. Even more challenging, the Mach numbers derived from X-ray observations for several relics have been found to be very low, too low to explain the relic luminosity by acceleration of electrons from the thermal pool~\cite{2020A&A...634A..64B}. It has been proposed that re-acceleration of pre-existing mildly relativistic electrons would reconcile acceleration at the shock front with the observed luminosities and the overall spectrum~\cite{2016ApJ...823...13K}. However, X-ray estimates also tend to be lower than the actual shock strength due to projection effects~\cite{2021MNRAS.506..396W}. Even if the origin of the high luminosities and the spectral indices is not fully understood, it is without doubt that magnetic fields are present in the cluster outskirts, electrons get accelerated to relativistic energies at the shock front and loose their energy mainly via synchrotron and inverse Compton emission while the shock front continues to propagate outwards.  
 
 %
 Radio relics have been found to be considerably polarised, see~\cite{2019MNRAS.490.3987W} for a compilation. The polarisation properties vary from relic to relic, however, some of the properties are highly remarkable. The average fractional polarisation of radio relic emission reaches values up 30\,\%, indicating a surprisingly ordered magnetic field distribution in the emission region.  Some relics show locally even higher polarisation fractions, up to about 60\,\% see e.g.~\cite{2012A&A...546A.124V,2009A&A...494..429B,2010Sci...330..347V,2014ApJ...786...49L}. Some relics show a strong frequency dependent depolarisation which can be attributed to the  Faraday depth dispersion due to the magnetoionic intra-cluster medium (ICM) along the line-of-sight from the observer to the emission region. The depolarisation is affected by the amount of ICM along the line-of-sight, see e.g. the discussion in~\cite{2021arXiv210915237R}.  When correcting for the Faraday rotation along the line-of-sight one finds the intrinsic polarisation angle of the synchrotron emission. For several relics the magnetic field which corresponds to polarisation direction (i.e., the so-called polarisation B-vector) is astonishingly well aligned with the relic morphology, i.e., the orientation of the shock surface, most spectacularly for the Sausage relic~\cite{2010Sci...330..347V,2021ApJ...911....3D}. Other relics which show this alignment are for instance the Toothbrush~\cite{2012A&A...546A.124V}, the northern relic in PSZ1\,G096.89 + 24.17~\cite{2014MNRAS.444.3130D} and the relic in MACS J0717.5 + 3745~\cite{2021arXiv210915237R}. The orientation of the polarisation has revealed that the magnetic field is well aligned with the surface of the shock. 
 
 %
 The high degree of polarisation in radio relics may indicate a very homogeneous magnetic field in the emission region. In this scenario, the alignment of the polarisation with the shock surface would imply that the pre-shock magnetic field is already reasonably well aligned with the shock surface even before the shock is crossing.  Since most relics are tangential to the X-ray isophotes of clusters the magnetic field would have to be tangential as well. Current numerical simulations, addressing the origin of magnetic fields in the ICM, seem to predict a rather random magnetic field orientation~\cite{2019MNRAS.490.3987W}.  Alternatively to a homogeneous magnetic field, the compression of a small-scale tangled isotropically oriented field due to the shock may explain the polarisation properties~\cite{1998A&A...332..395E}. The compression amplifies the magnetic field components in plane of the shock, causing an anisotropic magnetic field distribution with the major axis of the distribution in the plane of the shock. This has been demonstrated in recent simulations where the polarisation of a shock propagating through a tangled magnetic field is studied~\cite{2021MNRAS.507.2714D}. To date, it remains unclear if the relic polarisation is dominated by a large-scale field, i.e., the magnetic field orientation is rather uniform in the emission region, or if the polarisation is dominated by a small-scale field, i.e., the emission is random but anisotropic in the emission region.

 %
 For the Sausage relic, evidence has been found that the intrinsic polarisation decreases downstream with distance to the shock front~\cite{2021ApJ...911....3D}. 
 There are several effects which may affect the observed profile. Firstly, the ageing of the electron energy distribution may make the energy distribution effectively more mono-energetic, leading to an increased fractional polarisation. Secondly, in the observed relics, there might be projection effects --which also affect the spectral index profile--, causing a contribution of emission inclined to the line-of-sight and, therefore, reducing the fractional polarisation  downstream~\cite{2021ApJ...911....3D}. Thirdly, turbulent motions in the ICM may keep changing the magnetic field distribution in the downstream region and may reduce the anisotropy in the magnetic field distribution~\cite{2021MNRAS.507.2714D}.

 %
 In this article we report on a toy numerical model which allows us to study the polarisation properties in the small-scale tangled field scenario. We describe in brief our model calculations and discuss in particular how the polarisation properties depend on the magnetic field strength.  
 
 \section{MoCaRePo: A Monte Carlo Approach for Computing Radio Relic Polarisation} 

 %
 Polarised synchrotron emission originates from sources with an anisotropic magnetic field distribution. For instance, a source with a homogeneous magnetic field results in highly polarised emission. In more complex situations, the emission weighted field distribution determines the polarisation properties. Following the scenario described in~\cite{1998A&A...332..395E}, we implemented a model in which the anisotropic field distribution is caused by compression due to shock fronts induced by galaxy cluster mergers. To compute the emission in the downstream region of the shock, we represent the energy distribution of the relativistic electrons by a large number of monoenergetic CRe bunches, integrate the energy losses while advected with the downstream plasma for each bunch, and compute the polarised emission from the time of injection until the CRe bunch energy is too low to cause synchrotron emission in the relevant frequency regime.

 %
 Firstly, we summarise the assumptions for designing the model. A fundamental ingredient for our computation is the assumption that we observe emission from a large number of magnetic orientations and strengths along the line-of-sight, more precisely, within the beam solid angle of the observation. Therefore, within the beam there is a field distribution causing the polarisation properties. We assume that CRe get injected at a  shock front as a result of DSA~\cite{1987PhR...154....1B} and that the CRe injection energy distribution is described by a power law with slope $s$. Subsequently, the CRe advect with the downstream plasma and suffer synchrotron, inverse Compton and Coulomb losses. Moreover, we assume that CRe do not move relative to the plasma of injection, allowing us to neglect processes as CRe diffusion. We assume that the time scales for pitch angle scattering are small compared to the cooling times, hence, we adopt pitch angle averaged loss rates. Finally, we assume that shock properties, e.g. its strength, change slower than the cooling times. In this quasi-stationary situation, summing up the time evolution represents the entire emission of the downstream region at a single time. Here, we compute the polarisation properties of a planar shock front observed edge on. We do not include effects due to a possible spherical geometry of cluster merger shocks.

 The key properties for determining the synchrotron emission are the upstream magnetic field, the compression at the shock front, the injection CRe energy spectrum, and the downstream cooling. To compute the resulting polarisation properties, we choose a simple Monte Carlo approach. More precisely, we compute the time evolution of monoenergetic CRe bunches. We start the evolution of the CRe bunches with the injection at the shock front. We choose the energy distribution of bunches and the number of CRe per bunch to represent the injection energy spectrum. We compute the evolution for $N$ monoenergetic CRe bunches. The energy $E_i$ of bunch $i$ is given by

 \begin{equation}
     \log E_i 
     = 
     \log E_{\rm min} + i \cdot \frac{ \log E_{\rm max} - \log E_{\rm min} } { N }
     ,
 \end{equation}
 where $E_{\rm min}$ and $E_{\rm max}$ are the minimum and maximum energy range considered in our modelling. The number of CRe in each of the electron bunches $i$ is assumed to be proportional to $10^{-s-1}$. Typically, we choose $10^4$ to $10^5$ bunches for our computation.

 %

 We adopt a magnetic field probability distribution and assign to each CRe a field strength and direction according to this distribution. We assume that the magnetic field upstream to the shock is isotropically distributed, that is, the probability to find a magnetic field vector in the direction $(\theta,\phi)$ amounts to 
 \begin{equation}
     {\rm d} {\cal P} (\theta,\phi) 
     = 
     \frac{1}{4 \pi} \sin \theta \, {\rm d} \theta \, {\rm d} \phi,
 \end{equation}
 where $\theta$ is the angle enclosed by the magnetic field and the shock normal, and $\phi$ the angle enclosed by the magnetic field and the plane defined by the shock normal and the line-of-sight. The magnetic field gets compressed at the shock front according to the jump conditions for a magnetised medium. This leads to an anisotropic magnetic field distribution, as demonstrated in recent simulations~\cite{2021MNRAS.507.2714D}.  The shock compression keeps the strength of the field component parallel to the shock normal unchanged. Most importantly, the compression amplifies the field component in the shock plane. Here, we apply the jump conditions for a magnetised shock. For a parallel shock, that is, where the magnetic field is parallel to the shock normal, the shock does not alter the magnetic field. For a perpendicular shock, the field is in the plane of the shock and
 its strength gets amplified by the compression ratio $r$ of the shock 
 \begin{equation}
     B_{\perp,\rm down} 
     = 
     r 
     B_{\perp,\rm up} \, ,
 \end{equation}
 where $B_{\perp,\rm up}$ is the upstream magnetic field component perpendicular to the shock normal and $B_{\perp,\rm down}$ is the downstream one. In case of oblique shocks and strong magnetic fields, the amplification depends on the pre-shock obliquity, see e.g.,~\cite{2017MNRAS.464.4448W} for a discussion. Since we average over an isotropic magnetic field distribution, we simplify the computation and assume an amplification of the tangential field component equal to the compression ratio for any field direction. We adopt here a polytropic index of 5/3 for the plasma resulting in a maximum compression ratio of four.

 %
 For each electron bunch we compute energy losses due to synchrotron emission, inverse Compton emission and Coulomb scattering.  We adopt the loss rates $b( E, B, n_{\rm e}, z )$ given in~\cite{1999ApJ...520..529S}
 \begin{equation}
   \frac{{ \rm d } E }{{ \rm d } t }
   =
   - m_{\rm e} c^2 \, b( E, B, n_{\rm e}, z ) 
     ,  
 \end{equation}
 where $n_e$ is the density of electrons, $c$ the speed of light and $z$ the redshift.

 The polarised synchrotron emission is computed according to the geometry determined by the inclination of the shock front with respect to the line-of-sight and the direction of magnetic field, see e.g.,~\cite{1985rpa..book.....R} for a detailed discussion. The quasi-stationary shock scenario allows us to compute the total synchrotron emission of the CRe downstream of the shock as the sum of the entire cooling process. The resulting overall polarisation fraction is affected by the dependence of the radio luminosity on the magnetic field amplification of the shock. In~\cite{2007MNRAS.375...77H} we derived that the emission of the entire downstream region is proportional to
 \begin{equation}
     \epsilon
     \propto 
     \frac{B^{1+\frac{s}{2}}}{B_{\rm CMB}^2 + B^2 } 
     \, ,
     \label{eq::Bfact}
 \end{equation}
 where $\epsilon$ is the emissivity of the shock area and $B_{\rm CMB}$ the magnetic field strength equivalent to the energy density of photons in the Cosmic Microwave Background (CMB). The shock compression makes the magnetic field distribution anisotropic, which also affects the radio emissivity and, therefore, the polarisation fraction as well.  Eq.\,\ref{eq::Bfact} shows that the ratio of $B$ to $B_{\rm CMB}$ and the slope of the electron energy distribution play a crucial role. Here, we have set up a parameter study to especially investigate the impact of these two properties. 
 
 \section{Results}

\begin{figure}[H]
 \includegraphics[width=0.36\textwidth]{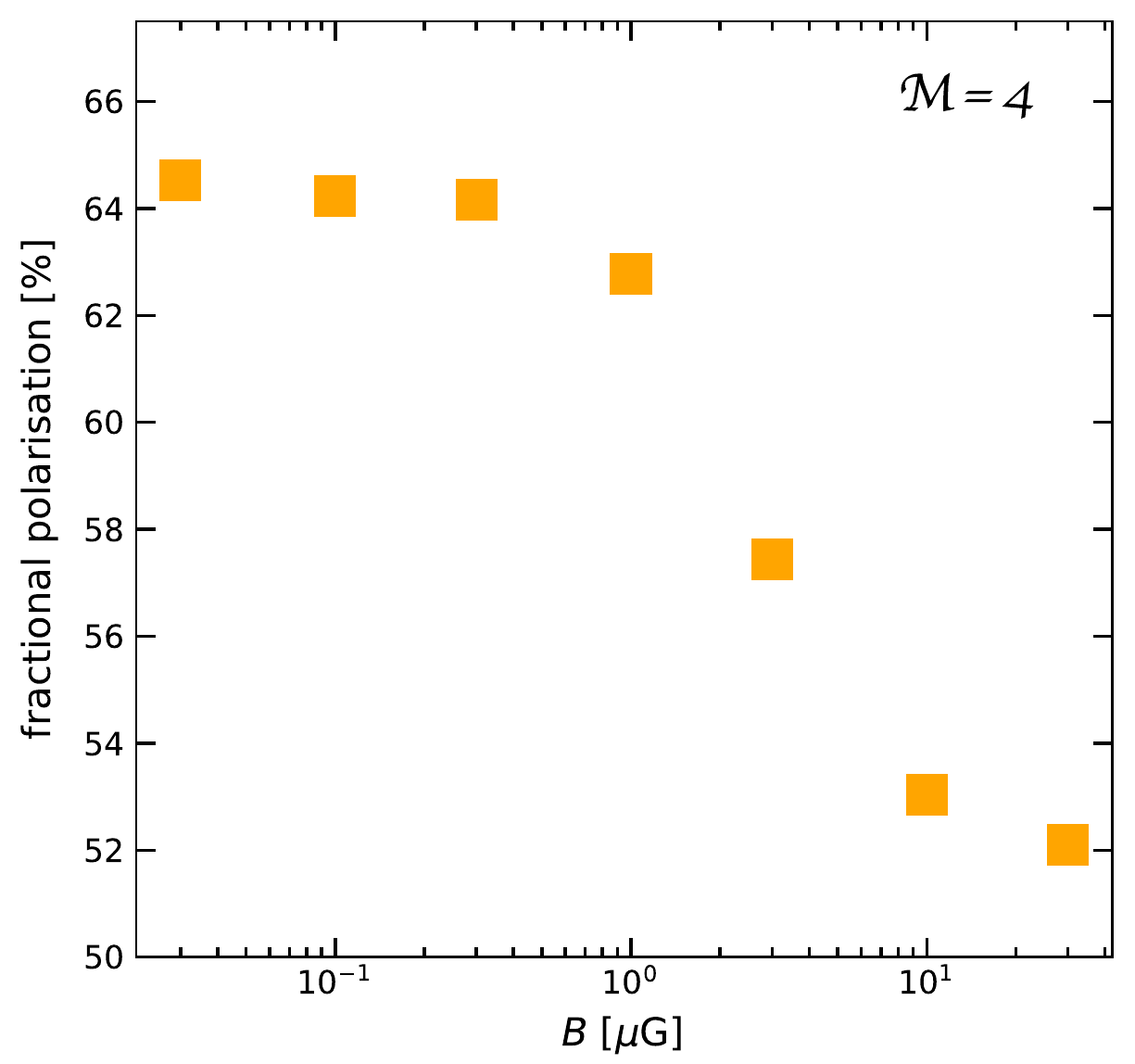}
 \hfill 
 \includegraphics[width=0.36\textwidth]{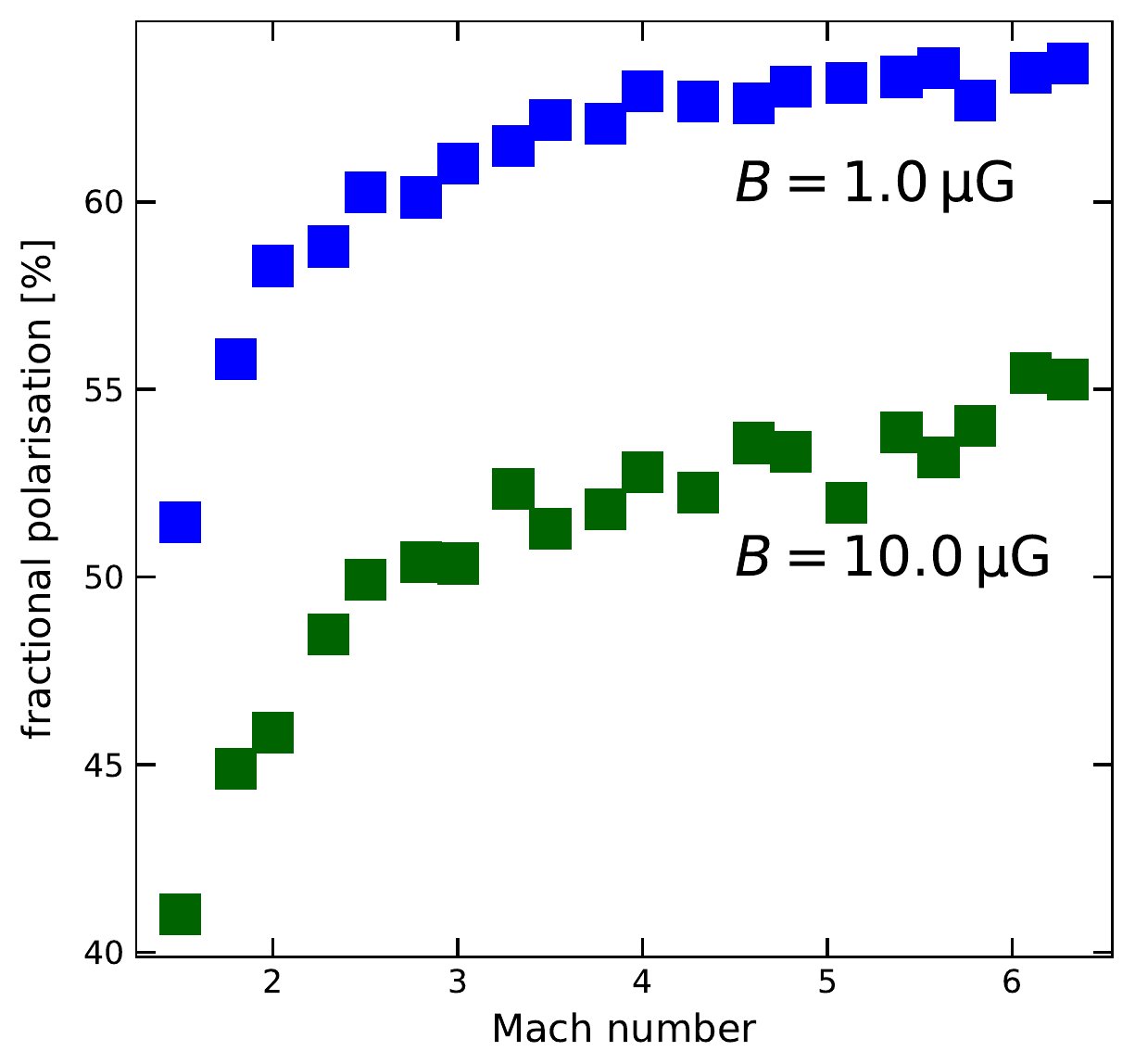}
 \caption{
   \textit{Left panel:} 
   Average fractional polarisation of the model relic as a function of magnetic field strength. The emission has been computed for a Mach number ${\cal M} = 4$ and a redshift of $z=0.1$. The fractional polarisation gives the luminosity weighted average including the entire downstream emission. For low magnetic fields ($B\ll B_{\rm CMB}$) the fractional polarisation is about 12\,\% higher than for high magnetic fields ($B\gg B_{\rm CMB}$). The scatter of data points is due to the Monte Carlo approach of the simulation.
   \textit{Right panel:}
   The fractional polarisation as a function of Mach number. The analysis includes the dependence of the emissivity on the Mach number beside the compression of the magnetic field at the shock front. The fractional polarisation has been computed for an upstream magnetic field strength of $B=1\,\upmu G$ (blue) and $B=10\,\upmu G$ (green). 
 \label{fig::B}}
 \end{figure}

 %
 As a first experiment, we investigate how the fractional polarisation depends on the magnetic field strength for a given shock strength. As shown in Figure~\ref{fig::B}, we find for a Mach number ${\cal M}=4$ that at very low magnetic field strengths the fractional polarisation obtained by  shock compression is up 65\,\%. For larger magnetic field strengths, that is for $B \gg B_{\rm CMB}$, the fractional polarisation is lower. For $B=30\,\upmu{\rm G}$ we find a fractional polarisation of 52\,\%. The dependence of the fractional polarisation on the magnetic field is plausible since for large magnetic fields, that is $B \gg B_{\rm CMB}$, the emissivity increases only slightly with the magnetic field, see Equation~\eqref{eq::Bfact}. The differences between the fractional polarisations for low  and for high magnetic fields is about 12\,\%. This is not large, however, it illustrates the transition between the two regimes. For $B \ll B_{\rm CMB}$ the amplification of the magnetic field component tangential to the magnetic field is accompanied by a strong increase in emissivity. For the $B \gg B_{\rm CMB}$ regime this is not the case, the emissivity depends only slightly on the field strength.
 
 %
 In the adopted scenario, also the strength of the shock affects the how the emissivity depends on the magnetic field, via the slope $s$. Observationally, it has been found that the polarisation fraction increases substantially with Mach number~\cite{2017A&A...600A..18K}. The reason for this dependence is that the compression factor (which affects the slope $s$) is higher for higher Mach numbers. However, the dependence of the emissivity on the magnetic field plays a crucial role and needs to be considered as well. In our second experiment, we investigate how the fractional polarisation depends on the Mach number. Since the slope $s$ decreases with higher Mach number, the emissivity has a smaller dependence on the magnetic field for stronger shocks, therefore the larger compression is to some extend compensated by the weaker dependence of the emissivity on the field strength. Therefore, also weak shocks are similarly high polarised as strong ones, in the small-scale tangled field compression model, see Figure~\ref{fig::B} (right panel).

 %
 As shown in Figure~\ref{fig::B} (right panel) and discussed before, the increase of the field strength from $1\,\upmu \rm G$ (blue) to $10\,\upmu \rm G$ (green) lowers the fractional polarisation by about 10\,\%. For the Mach numbers investigated here and all magnetic field strengths the polarisation fraction is significantly above the average polarisation found for radio relics. However, only a small margin for beam and Faraday rotation depolarisation would remain to reduce the expected fractional polarisation to the observed one in case of strong magnetic fields and a low Mach number. 
 
\begin{figure}[H]
 \begin{center}
 \includegraphics[width=0.5\textwidth]{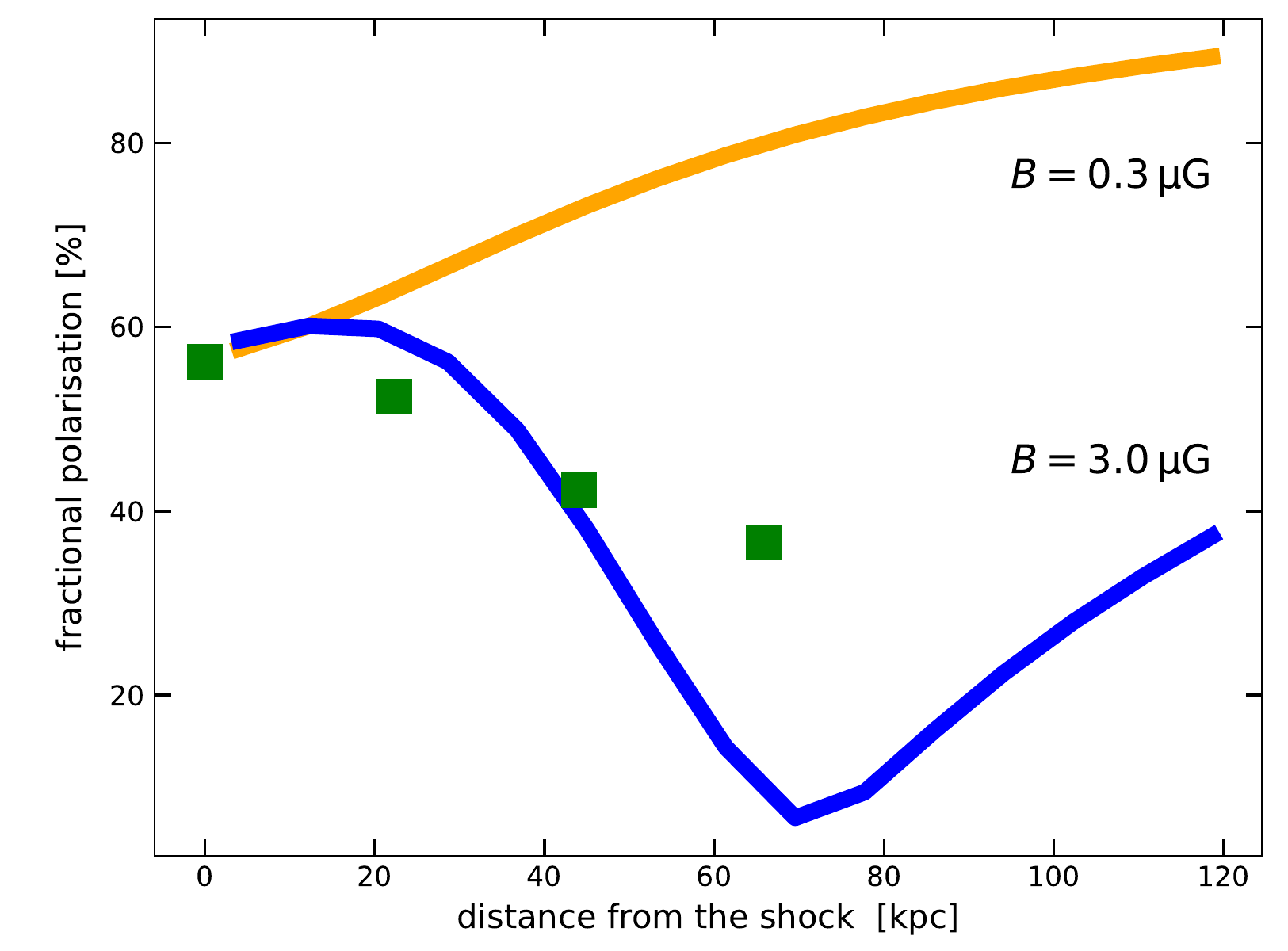}
 \caption{
   The downstream profile of the fractional polarisation.  We investigate the profiles for a Mach number ${\cal M}=4$ and two magnetic field strengths, namely $B=0.3\,\upmu \rm G$ (orange) and $B=3\,\upmu \rm G$ (blue). The downstream polarisation profile has a minimum at about 70\,kpc, where we have adopted a shock propagation speed of ${1000\,\rm km\,s^{-1}}$. The increase in polarisation at distances larger than 70\,kpc for a magnetic field strength $B=3.0\,\upmu {\rm G}$ corresponds to a very low surface brightness. 
   We also plot the downstream fractional polarisation of the Sausage relic as 
   reported in \cite{2021ApJ...911....3D}. The decrease of downstream polarisation in the Sausage relic is less prominent than in our model, this may for example reflect a mixture of magnetic field strengths in the downstream region.  
   }
 \label{fig::downstream}
 \end{center}
 \end{figure}

 %
 Finally, we investigate the fractional polarisation as a function of distance to the shock front, that is, the downstream polarisation fraction profile. As shown by~\cite{2021MNRAS.507.2714D}, the fractional polarisation increases downstream of the shock for a homogeneous magnetic field. This effect originates from the electron spectrum ageing. Effectively, the slope steepens in that part of the electron spectrum which is relevant for the observed synchrotron emission. A steeper electron energy distribution results in a higher fractional polarisation. We study the downstream fractional polarisation profile using the MoCaRePo tool, see Figure~\ref{fig::downstream}. In contrast to an homogeneous field, an idealised situation which has been included in the~\cite{2021MNRAS.507.2714D}, we consider here a shock compressed field distribution, not a homogeneous one, hence the polarisation fractions are lower. However, the fractional polarisation increases downstream for a low magnetic field (orange line). 
 
 %
 Interestingly, the downstream fraction polarisation behaves quite differently for a higher magnetic field strength. Here, we investigate the profile for a field strength $B=3\,\mu\rm G$, that is, a field strength believed to be possibly realised in radio relics. We find that the fractional polarisation decreases in the downstream regime. We speculate that the reason is as follows: CRe bunches with a quasi-perpendicular magnetic field orientation, that is the magnetic field vector is basically parallel to the shock plane, experience the strongest amplification of the field. For these bunches the cooling is shorter than for bunches with a quasi-parallel field orientation. We do find that just at the shock front the fractional polarisation is about 60\,\%, that is, it is consistent with highest fractional polarisations found in relics. Further downstream the CRe energy distribution in bunches with a quasi-perpendicular field orientation has significantly cooled, therefore, these bunches loose radio luminosity faster than the bunches with a quasi-parallel field. {In Figure~\ref{fig::downstream}, we also show the downstream polarisation measured for the Sausage relic~\cite{2021ApJ...911....3D}. Evidently, the decrease in downstream polarisation is less prominent in the Sausage relic as compared to our model. However, a simple explanation for that would be mixture of magnetic field strengths in the Sausage relic. } Since the energy of the electrons which dominate the synchrotron emission depends on the observing frequency, the effect discussed here would be frequency dependent. This implies, that intrinsic fractional polarisation of radio relics, as a function of distance to the shock front, also depends on the observing frequency. A more detailed investigation is necessary to investigate the possible impact of that effect on the observed fractional polarisation profiles.

 It should be noted that these simple toy model experiments are based on many simplifying assumptions, for instance: there is a plane shock front, there are no projection effects, no diffusion takes place, there is fast pitch angle isotropisation, there is an isotropic magnetic field distribution upstream, and the absolute magnetic field strength is uniform, that is there are no small regions with a high B which may dominate the emission. It is beyond the scope of this work to investigate how appropriate these assumptions are for a realistic radio relic formation scenario. The main focus of this work was to illustrate that magnetic field strength, more precisely the ratio between magnetic field strength and equivalent field strength of the Cosmic Microwave Background, may crucially affect the fractional polarisation properties of radio relics.

\section{Discussion and Conclusion}

 %
 Using our toy model MoCaRePo for relic polarisation, we demonstrated that shock compression can lead to polarisation fractions of 50\,\% and above. This underlines that the shock compression scenario provides a plausible explanation for the polarisation of radio relics, most importantly, because it naturally explains the observed alignment of the polarisation orientation (B-vector) with the shock surface. However, more sophisticated numerical studies are required to investigate in detail the polarisation caused by a shock front propagating into a medium containing a tangled magnetic field. Current simulations~\cite{2021MNRAS.507.2714D} basically support the shock compression scenario. They also provide evidence that the polarisation direction may significantly fluctuate at the shock front and, therefore, the averaged polarisation fraction might be lower. However, the polarisation fractions averaged over a larger area are still up to about 50\,\%. It should be noted that in particular for the field with turbulence driven on the small scales ($L/4$-model in~\cite{2021MNRAS.507.2714D}) the fractional polarisation decreases quickly with increasing beam solid angle and seems to be too low compared to fractional polarisation found the Sausage relic and the relics in Abell\,2744. In contrast, the field obtained from turbulence driven on larger scales ($2L/3$) shows a higher fractional polarisation even when averaged with large beams sizes. 
 %
 
 %
 The polarisation of radio relics provides an excellent opportunity to investigate the strength and morphology of magnetic fields in galaxy clusters. However, these studies would benefit from knowing clearly if the polarisation of radio relics is a result of a compressed small-scale tangled field or if it rather reflects the large scale morphology of the magnetic field. The work presented here highlights that we may expect in the compressed small-scale tangled field a decreasing downstream fractional polarisation profile, which is up to now not explained if the polarisation is a result of a large scale field. Investigating the polarisation profile of radio relics is a crucial contribution to unriddle the origin of radio relic polarisation.


 
 We introduced an additional consideration for understanding the polarisation profiles of radio relics. It has been shown in recent simulation~\cite{2021MNRAS.507.2714D} that for a shock front propagating into a tangled downstream medium a decrease in polarisation is expected downstream. The simulations provided evidence that the compression at the shock front causes an anisotropy in the magnetic field distribution, which apparently returns to a more isotropic distribution further downstream. A more detailed analysis is necessary to investigate to which extent the decrease in polarisation found in~\cite{2021MNRAS.507.2714D} comprises the dependence of the cooling time on the magnetic field discussed here. Both effects, turbulent motions in the downstream plasma and different cooling times may contribute to the decrease in fractional polarisation downstream of the shock front. 
 
 We introduced here the tool MoCaRePo which is deliberately designed to permit a simple incorporation of additional effects. For instance, the effect of downstream turbulence can be added by slowly rotating the downstream magnetic field vectors. This would cause an even stronger reduction of the downstream polarisation. We plan to incorporate several of these effects and to compare to the observed polarisation profiles in an upcoming work.

 To summarise, we demonstrated that the compression of a small-scale tangled magnetic field provides a plausible scenario for the origin of the polarisation of radio relics. Upcoming high resolution studies of the polarisation in radio relics will shed light on the relic polarisation and the magnetisation of the ICM.

\vspace{6pt} 



\authorcontributions{M.H. conceptualised and implemented the MoCaRePo code and performed the simulation runs. M.H., K.R., D.W., G.d.G. P.D.-F. analysed and interpreted the data. All authors have read and agreed to the published version of the manuscript.}

\funding{M.H. acknowledges support by the BMBF under the grant number 05A20STA. KR acknowledges financial support from the ERC Starting Grant ``MAGCOW" no. 714196. D.W. is funded by the Deutsche Forschungsgemeinschaft (DFG, German Research 227 Foundation) - 441694982. P.D.F was supported by the National Research Foundation (NRF) of Korea through grants 2016R1A5A1013277 and 2020R1A2C2102800.}

\institutionalreview{Not applicable}

\informedconsent{Not applicable}


\dataavailability{We are happy to share the MoCaRePo results presented here upon request.}

\conflictsofinterest{The authors declare no conflict of interest. The funders had no role in the design of the study; in the collection, analyses or interpretation of the data; in the writing of the manuscript, or in the decision to publish in this work.} 




\begin{adjustwidth}{0cm}{0cm}

\reftitle{References}

\end{adjustwidth}
\end{paracol} 
\end{document}